\documentclass[12pt]{iopart}

\usepackage{iopams}
\usepackage{bbm,wasysym}
\usepackage{srcltx}
\usepackage{graphicx}
\usepackage{verbatim}
\newcommand{\id}{{\mathbbm{1}}}

\newcommand{\bra}[1]{{\langle #1 |}}
\newcommand{\ket}[1]{{| #1 \rangle}}
\newcommand{\bracket}[2]{{\langle #1 | #2 \rangle}}

\begin{document}

\title[Optimal super dense coding over noisy quantum channels]{Optimal super dense coding over noisy quantum channels}

\author{Z  Shadman$^1$ $^\ast$, H  Kampermann$^1$, C  Macchiavello$^2$ and D  Bru\ss$^1$}

\address{$^1$ Institute f\"ur Theoretische Physik III,
              Heinrich-Heine-Universit\"at
              D\"usseldorf, D-40225 D\"usseldorf, Germany}
  \address{$^2$ Dipartimento di Fisica  ``A. Volta" and INFM-Unit$\acute{a}$ di Pavia, Via Bassi 6, 27100, Pavia, Italy }
\ead{$^\ast$ shadman@thphy.uni-duesseldorf.de}

\begin{abstract}
We investigate super dense coding in the presence of noise, i.e.,
the subsystems of the entangled resource state have to pass a 
noisy unital quantum channel between 
the sender and the receiver.
We discuss explicitly the case of Pauli channels in arbitrary dimension and
derive the super dense coding capacity (i.e. the optimal information transfer) 
for some  given resource states. For the qubit
depolarizing channel, we also optimize the super dense coding capacity with respect to the input state.  We show that below a threshold value of the noise parameter the super dense coding protocol is
optimized by a maximally entangled  initial state, while above the 
threshold it is optimized by a product state.
Finally, we
provide an example of a noisy channel where non-unitary pre-processing
increases the super dense coding capacity, as compared to only unitary encoding.

\end{abstract}

\pacs{03.67.-a, 03.67.Hk, 03.65.Ud}
\maketitle

\section{Introduction}
In quantum information processing, entanglement can be used as a resource for super dense
coding, as
introduced by Bennett et al. \cite{Bennett}. 
Essential  to this communication protocol is an entangled initial state that is shared between sender(s) and receiver(s), together with the property that an entangled state can be transformed 
by the sender into another state 
via a \emph{local} operation, taken from some set of operations.
The sender's subsystem is then transmitted to the receiver (ideally via a noiseless
channel), who identifies the global state in an optimal way. 
The super dense coding capacity is defined to be the maximal amount of classical information that 
can be reliably transmitted to the receiver for a given initial state. In the last years attention has 
been given to various scenarios of super dense coding over noiseless channels \cite{hiroshima, ourPRL, Dagmar}.
It has been  proved that for  noiseless channels and for unitary encoding,
the super dense coding capacity is given  by \cite{hiroshima}
\begin{eqnarray}
C=\log d+S(\rho_b)-S(\rho)\;,
\label{Cnoiseless}
\end{eqnarray}
where $\rho$ is the initial resource state shared between the sender (Alice) and 
the receiver (Bob). Here, 
$d$ is the dimension of Alice's system,
$\rho_b$ is Bob's reduced density operator and $S(\rho)=-\tr(\rho \log \rho) $ is 
the von Neumann entropy. Without the additional resource of entangled states, 
a $d$-dimensional quantum state can be used to transmit 
the information  $\log d$. Hence, quantum states  for which $S(\rho_b)-S(\rho)>0 $ are the  states 
which are useful for dense coding. The relation $S(\rho_b)-S(\rho)>0 $ cannot hold for quantum states with positive partial transpose
\cite{ourPRL}.
Therefore, states that are useful for dense coding always have a
non-positive partial transpose (NPT). However, the converse is not true: There exist states 
which are NPT but which are not useful for dense coding. 
One can then classify bipartite states according to their usefulness for
super dense coding \cite{Dagmar}. 
Besides the case of a single sender and receiver sharing an initial pure entangled state
and using unitary encoding some other
scenarios also have been discussed: many senders and either one or two receivers, 
 initially entangled mixed states, non-unitary encoding, etc. \cite{Bennett, hiroshima, Dagmar, CPTP}.
Super dense coding has been realized
 in optical experiments with polarized photons by Mattle et al. \cite{Zeilinger}, 
and for continuous variables by Li \etal \cite{Peng}. 

In a realistic scenario however noise is unavoidably present. 
The central theme of this paper is the question: how does noise in the transmission
channel affect the superdense coding capacity? Here, we focus on the case of
a single
sender and a single receiver, assuming unitary encoding at first, and then
generalizing to non-unitary encoding.
Physically,  noise is a process that arises through  interaction 
 with the environment. 
Mathematically, a noisy quantum channel can be described  as a completely positive trace 
preserving linear map $\Lambda$, acting  on the quantum state.  In this paper we 
will study two different scenarios of noisy channels:
first, we will assume that the sender Alice and the receiver Bob share already a 
bipartite quantum state $\rho$ (it could e.g. have been distributed to them by
 a third party). After Alice's local encoding operation, she sends
her part of the quantum state to Bob via the noisy channel, described by the
map $\Lambda_a$, see Figure 1. We call this the case of a \emph{one-sided}
channel. Second, we consider the case where Alice prepares the bipartite state
$\rho$ and sends one part of it via a noisy channel, described by the map 
$\Lambda_b$, to Bob, thus establishing the shared resource state for super dense coding.
When the two parties want to use this resource, Alice does the local encoding and
then sends her part of the state via the channel $\Lambda_a$ to Bob, see Figure 2. We call
this case a \emph{two-sided} channel.

\begin{center}
\begin{tabular}{c  }
\includegraphics[width=8cm]{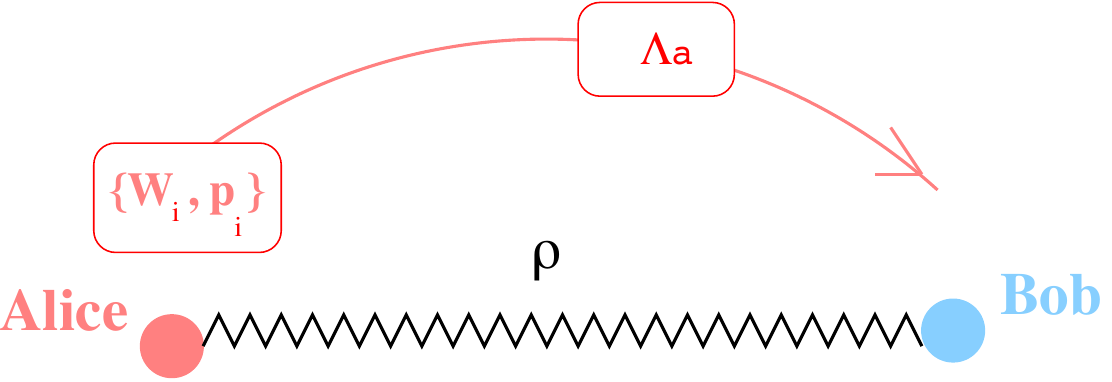}  \\
\end{tabular}
\end{center}

\vspace{0.5cm}

Figure 1. { \scriptsize \emph{One-sided} noise: Bipartite super dense coding  with an  initially entangled state $\rho $,
shared between Alice and Bob.  Alice applies the  unitary operator   $W_i$,
taken from a set $\{W_i\}$ with probability $\{p_i\}$,  on her part of  the entangled 
state $\rho $.  She  sends the encoded state with probability $p_i$ over a noisy channel, described by
the map $\Lambda_a$,  to Bob.  In the first approach we assume that   $\Lambda_a$  just affects 
 Alice's subsystem, but  that there is no noise  on Bob's side. }

\begin{center}
\begin{tabular}{c  }
\includegraphics[width=8cm]{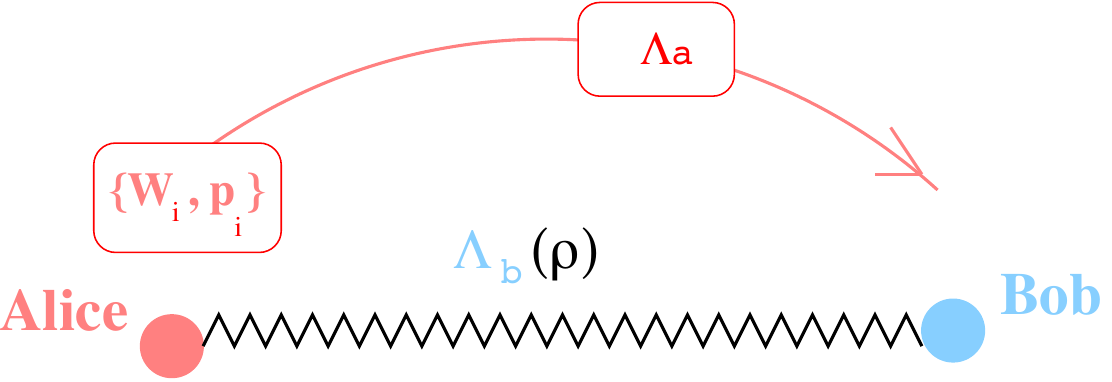}  \\
\end{tabular}
\end{center}

\vspace{0.5cm}

Figure 2. { \scriptsize \emph{Two-sided} noise:
Bipartite super dense coding  with an  initially entangled state $\rho $, shared
 between Alice and Bob. In the second approach,  the  noisy channel  $\Lambda_a$
  influences   Alice's subsystem after encoding   while the noisy channel  $\Lambda_b$ 
 has already affected  Bob's side in the distribution step of the initial
state $\rho$.}

\vspace{1cm}

The paper is organized as follows: in Section II we discuss
the definition of the
Holevo quantity for an ensemble of states in the presence of a noisy channel.
 We introduce a certain condition on the von Neumann entropy and we derive the
super dense coding capacity for those cases where this condition is fulfilled.
In Sections III and IV, we give examples of initial states and channels for
which this  condition on the von Neumann entropy is satisfied, and 
calculate their optimal super dense coding capacity explicitly.
Section V provides a comparison between the super dense coding capacities in the 
presence of a \emph{one-sided} or \emph{two-sided} 2-dimensional depolarizing channel, 
and the classical capacity of a 2-dimensional depolarizing channel.
In Section VI we consider the case of non-unitary encoding and show an example
where pre-processing is useful to increase the dense coding capacity of the
initial resource  state in the presence of  the noisy channel.

\section{Super dense coding capacity}

In the super dense coding protocol  Alice and Bob share a bipartite
entangled quantum state $\rho$.  Alice performs local unitary operations $W_i$ with
probability $p_i$ (where $ \sum_{i}p_i=1$) on $\rho$  to encode classical
information through the state $\rho_i$, i.e. 
\begin{eqnarray}
 \rho_i= (W_i\otimes\id)\rho({W_i}^{\dagger}\otimes\id).
\label{optimalunitaryencoding}
\end{eqnarray}
We consider $\Lambda:\rho_i \rightarrow \Lambda(\rho_i)$ to be any completely
positive map that acts on the shared state $\rho_i$. 
(Below $\Lambda$ will describe the noise acting on the ensemble states.)
The ensemble that Bob(s)
receives is $\{\Lambda(p_i,\rho_i)\}$. 
The amount of classical information transmitted via a quantum channel is measured by the 
Holevo quantity or $\chi$-quantity. This quantity for the ensemble $ \{\Lambda(p_i,\rho_i)\}$ is given by
\begin{eqnarray}
\chi=S\left(\overline{\Lambda(\rho)}\right)-\sum_i p_i S\left(\Lambda(\rho_i)\right)
=\sum_{i}p_iS\left(\Lambda(\rho_i)\|\overline{\Lambda(\rho)}\right),
\label{Holevo}
\end{eqnarray}
where $\overline{\Lambda(\rho)}=\sum_i p_i \Lambda(\rho_i)$ is the average
state and $ S(\eta)$ is the von Neumann entropy of
$\eta $. The symbol
$S(\sigma\|\rho)$ denotes the relative entropy, defined as
$S(\sigma\|\rho)=\tr(\sigma\log\sigma-\sigma\log\rho )$.
Note that $\chi$ is a function of the resource state $\rho$, the encoding
$\{p_i,W_i\}$ and the channel $\Lambda$. For brevity of notation we will  not write
explicitly these arguments of $\chi$.

The super dense coding capacity $C$ for a given resource state $\rho$
is defined to be the maximum of the  Holevo
quantity $\chi$ with respect to $\{p_i,W_i\}$, that is
\begin{eqnarray}
 C=\max_{\{p_i,W_i\}}(\chi).
\end{eqnarray}

In this paper we consider bipartite systems, where each subsystem has
 finite dimension $d$.
A general density matrix on $ \mathrm{C^d}\otimes \mathrm{C^d}$ in the Hilbert-Schmidt representation can be conveniently decomposed as

\begin{eqnarray}
\rho=\id\otimes\frac{\rho_b}{d}+\frac{1}{d^2}\left(\sum_{i=1}^{d^2-1} r_i\lambda_i\otimes\id+\sum_{i,k=1}^{d^2-1}t_{ik}\lambda_i\otimes\lambda_k\right),
\label{rho}
\end{eqnarray}
where $\rho_b=\tr_a \rho
$ represents Bob's reduced density operator and $\lambda_i$ 
are the generators of the $\mathrm{SU}(d)$ algebra with  $\tr\lambda_i=0$. The parameters $ r_i,s_i,t_{ik}$ are real numbers.
We introduce the set of unitary operators $\{V_i\}$, defined as

\begin{eqnarray}
V_{i=(m,n)}\ket j= \exp(\frac{2\pi \mathrm{i}nj}{d})\ket {j+m(\mathrm{mod}\,  d)}.
\label{1}
\end{eqnarray}
These operators satisfy the condition $d^{-1} \tr({V_i}{V_j^{\dagger}})=\delta_{ij}$. Integers $m$ and $n$ run from 0 to $d-1$ such that we have $d^2$ unitary operators $V_i$.
We will consider in the following the case of unital noisy channels acting on
Alice's and Bob's systems, namely channels described by the completely
positive map

\begin{eqnarray}
\Lambda(\rho)=\sum_{m}K_{m}\rho K^{\dagger}_{m}\;, \quad
\sum_{m} K^{\dagger}_{m}K_{m}=\id\;, \quad
\sum_{m} K_{m}K^{\dagger}_{m}=\id\;,
\label{unital}
\end{eqnarray}
where $K_m$ are Kraus operators.
Here,  the first  condition on the Kraus operators corresponds to 
trace preservation, and the second condition guarantees
the unital
property $\Lambda(\id)=\id$.
We will show in this section that for unital memoryless noisy quantum channels and 
certain
initial resource states, the set of unitary operators $\{V_i\}$ with equal probabilities
is the optimum encoding
and leads to the maximum of the Holevo quantity.

We will first prove in Lemma 1 some properties that hold for the specific
encoding $\{V_i\}$. In the following the symbol $\tau_i$ will denote
the resource state after encoding with $V_i$, whereas $\tau$ will 
denote the resource state after encoding with an arbitrary  unitary operation
$U$. The ensemble average after the specific encoding with $\{V_i\}$, the probability distribution  $p_i=1/d^2$
and after action of the channel will be denoted
as $\tilde{\rho}$. -
For similar methods in the case of noiseless channels see also \cite{hiroshima}.

{\bf Lemma 1.} Let $\Lambda_a(\sigma_a)=\sum_{m} A_{m}\sigma_a A_{m}^\dagger$
and  $\Lambda_b(\sigma_b)=\sum_{\tilde m} B_{\tilde m}\sigma_b
B_{\tilde m}^\dagger$ be any two unital channels  which act on Alice's
and Bob's side, respectively.
For an initial resource state $\rho$ shared between Alice and Bob, the global channel
$\Lambda_{ab}$ then acts as

 \begin{eqnarray}
\Lambda_{ab}(\rho)=\sum_{m,\tilde m} \left(A_{m}\otimes B_{\tilde m}\right)
\rho \left(A_{m}^\dagger\otimes B_{\tilde m}^\dagger\right).
\label{def-channel}
\end{eqnarray}
 Then, the following statements hold:\newline
{\bf 1-a)} For $ \tau_i=(V_i\otimes\id)\rho({V_i}^{\dagger}\otimes\id)$, with
$V_i$ being defined in (\ref{1}), the average $\tilde{\rho}$  of the
ensemble $\{p_i=\frac{1}{d^2},\Lambda_{ab}(\tau_i)\}_{i=0}^{d^2-1}$
takes the form $\tilde{\rho}=\id \otimes\Lambda_b(\frac{\rho_b}{d})$.\newline
{\bf 1-b)}
For  $\tau=\left(U\otimes\id\right)\rho\left(U^\dagger\otimes\id\right)$
with  $U$ being any unitary operator acting on Alice's system,
$\tr \left(\Lambda_{ab}(\tau)\log \tilde{\rho}\right)=-S(\tilde{\rho})$.\newline
{\bf 1-c)} The relative entropy between $\Lambda_{ab}(\tau)$ and $\tilde{\rho}$
can be expressed as
$S\left(\Lambda_{ab}(\tau)\|\tilde{\rho}\right)=S\left(\tilde{\rho}\right)-S\left(\Lambda_{ab}(\tau)\right)$.\newline
{\bf Proof 1-a).}
In \cite{hiroshima} it was shown that the average of the ensemble \newline
$\{ p_i=\frac{1}{d^2},\tau_i\}_{i=0}^{d^2-1}$ is
\begin{eqnarray}
\sum_i\frac{1}{d^2}\tau_i=\id\otimes   \frac{\rho_b}{d}.
\label{averagenoiseless}
\end{eqnarray}
By using (\ref{averagenoiseless}), the linearity of the channel and its
unital property, the average of the ensemble $\{p_i=\frac{1}{d^2},\Lambda_{ab}(\tau_i)\}_{i=0}^{d^2-1}$ is

\begin{eqnarray}
\tilde{\rho}&=&\sum_i\frac{1}{d^2}\Lambda_{ab}(\tau_i)
= \Lambda_{ab} (\frac{\id}{d}\otimes\rho_b)
=\id \otimes\Lambda_b(\frac{\rho_b}{d}).
\end{eqnarray}
{\bf Proof  1-b).}
In Lemma (1-a) we showed that $\tilde{\rho}=\id
\otimes\Lambda_b(\frac{\rho_b}{d})$ and hence,
$\log\tilde{\rho}=\id \otimes\log\Lambda_b(\frac{\rho_b}{d})$. Therefore:
\newpage

\begin{eqnarray}
&&\tr\left(\Lambda_{ab}(\tau)\log \tilde{\rho}\right)= \tr\left[\left(\sum_{m}A_{m}UU^\dagger A^\dagger_{m}\right)\otimes
\left(\sum_{\tilde m}B_{\tilde m}\frac{\rho_b}{d}
B_{\tilde m}^\dagger\log\Lambda_b(\frac
{\rho_b}{d})\right)\right.\nonumber\\
&&+\left.\frac{1}{d^2}\left(\sum_{i=1}^{d^2-1} r_i
\sum_m A_m U \lambda_i U^\dagger A_m^\dagger\right)\otimes
\left(\sum_{\tilde m}B_{\tilde m}
B_{\tilde m}^\dagger\log\Lambda_b(\frac
{\rho_b}{d})\right)\right.\nonumber\\
&&+\left.\frac{1}{d^2}\sum_{i,k=1}^{d^2-1} t_{ik}
\left(\sum_m A_m U \lambda_i U^\dagger A_m^\dagger\right)\otimes
\left(\sum_{\tilde m}B_{\tilde m} \lambda_k
B_{\tilde m}^\dagger\log\Lambda_b(\frac
{\rho_b}{d})\right)\right]\;.
\end{eqnarray}

By using the linearity of the trace and the relations

\begin{eqnarray}
\tr[\sum_{m}A_{m}UU^\dagger A^\dagger_{m}]=
\tr[\sum_{m}A_{m}A^\dagger_{m}]= \tr[\id]\;,
\end{eqnarray}

\begin{eqnarray}
\tr[\sum_{m}A_{m}U\lambda_i U^\dagger A^\dagger_{m}]&=&
\tr[U\lambda_iU^\dagger\sum_{m}A^\dagger_{m}A_{m}]
 \nonumber\\
&=&\tr[U\lambda_i U^\dagger]=\tr[\lambda_i] =0
\end{eqnarray}

we can write

\begin{eqnarray}
\tr\left(\Lambda_{ab}(\tau)\log \tilde{\rho}\right)&=&
\tr_a\tr_b\left[\sum_{m,\tilde m}
\id\otimes\left(B_{\tilde m}\frac{\rho_b}{d}
B_{\tilde m}^\dagger\log\Lambda_b(\frac{\rho_b}{d})\right)\right]\nonumber\\
&=&\tr_b\left[\Lambda_b(\rho_b)\log\Lambda_b(\frac{\rho_b}{d})\right]=
-S(\tilde{\rho}).
\end{eqnarray}
{\bf Proof 1-c).} Using the definition of the relative entropy $S(\sigma\|\rho)=\tr(\sigma\log\sigma-\sigma\log\rho )$ and the result of Lemma (1-b) we can
write

\begin{eqnarray}
S(\Lambda_{ab}(\tau)\|\tilde{\rho})&=&\tr(\Lambda_{ab}(\tau)\log\Lambda_{ab}(\tau)-\Lambda_{ab}(\tau)\log \tilde{\rho})\nonumber\\
&=&S(\tilde{\rho})-S(\Lambda_{ab}(\tau)).
\end{eqnarray}
 \hfill{$\Box$}

We now show that for resource states with a certain symmetry
property, namely for those states where the von Neumann entropy after
the channel action is independent of the unitary encoding, the encoding
with the equally probable operators $\{V_i\}$, as given in (\ref{1}), is optimal.
Our proof follows the line of argument developed in  \cite{hiroshima}.

{\bf Lemma 2.} Let $\tau_i$ denote the resource state after encoding
with $V_i$, given in (\ref{1}). Let
\begin{eqnarray}
 \tilde{\chi}= S(\tilde{\rho})-\frac{1}{d^2}\sum_{i}^{d^2-1}S(\Lambda_{ab}(\tau_i))
\label{holevoequal}
\end{eqnarray}
 be the Holevo
quantity for the ensemble $ \{p_i=\frac{1}{d^2},\Lambda_{ab}(\tau_i)\}$,
where $\tilde{\rho}$ is the average state of this ensemble and  $\Lambda_{ab}(\cdot)$
is defined in (\ref{def-channel}).  For all the channels $\Lambda_{ab}$ 
and all initial states $\rho$
for which
\begin{eqnarray}
S(\Lambda_{ab}(\tau))=\frac{1}{d^2}\sum_{i}^{d^2-1}S(\Lambda_{ab}(\tau_i))
\label{condition}
\end{eqnarray}
holds, $\tilde{\chi}$ is the super dense coding capacity.
Here    $\tau= \left(U\otimes\id\right)\rho\left(U^\dagger\otimes\id\right) $,
as we defined already above, with $U$ being any unitary operator.\newline
{\bf Proof.}
Let us consider  an arbitrary encoding, leading to an ensemble
$\{p_i,\Lambda_{ab}(\rho_i)\}$. We will show that its Holevo quantity $\chi$
cannot be higher than $\tilde{\chi}$ in (\ref{holevoequal}), if the condition
(\ref{condition}) is fulfilled.
 
 {If} $S(\Lambda_{ab}(\tau))=\frac{1}{d^2}\sum_{i}^{d^2-1}S(\Lambda_{ab}(\tau^i))$, then from  (\ref{holevoequal}) and Lemma (1-c),
\begin{eqnarray}
\tilde{\chi}= S(\Lambda_{ab}(\tau)\|\tilde{\rho}).
\end{eqnarray}
Since this equation holds for any $\tau$ that fulfills (\ref{condition}), it specially holds for $\rho_i$, i.e.
\begin{eqnarray}
\tilde{\chi}= S(\Lambda_{ab}(\rho_i)\|\tilde{\rho})=\sum_i p_i S(\Lambda_{ab}(\rho_i)\|\tilde{\rho}).
\end{eqnarray}
Using  Donald's identity, see \cite{Donald}, the right hand side of the above equation can be decomposed as
\begin{eqnarray}
\sum_i p_iS(\Lambda_{ab}(\rho_i)\|\tilde{\rho})=\sum_{i}p_iS(\Lambda_{ab}(\rho_i)\|\overline{\Lambda_{ab}(\rho)})
+S(\overline{\Lambda_{ab}(\rho)}\|\tilde{\rho})
\end{eqnarray}
with $\overline{\Lambda_{ab}(\rho)}= \sum_i p_i \Lambda_{ab}(\rho_i)$. The first term on the right hand side is the Holevo quantity for any arbitrary ensemble $\{p_i,\Lambda_{ab}(\rho_i)\}$. Hence,
\begin{eqnarray}
\tilde{\chi}=\chi+S(\overline{\Lambda_{ab}(\rho)}\|\tilde{\rho}).
\end{eqnarray}
Since the relative entropy $S(\overline{\Lambda_{ab}(\rho)}\|\tilde{\rho})$ is always 
positive or zero we can say that $\tilde{\chi}$ is always bigger or equal than $\chi$ and hence,  $\tilde{\chi}$ is the super dense coding capacity.
\hfill{$\Box$}

From Lemma 2 we find that 
\begin{eqnarray}
\tilde{\chi}=S(\tilde{\rho})-S(\Lambda_{ab}(\tau)).
\end{eqnarray}
Since the above equation holds for $\tau= \left(U\otimes\id\right)\rho\left(U^\dagger\otimes\id\right) $ with any unitary $U$, it especially holds for $\tau=\rho$. Hence, whenever the condition
(\ref{condition}) is true,  the super dense coding capacity is given by
\begin{eqnarray}
C=\tilde{\chi}=S(\tilde{\rho})-S(\Lambda_{ab}(\rho)),
\label{capacity}
\end{eqnarray}
where $\tilde{\rho}$ is the average of the ensemble after encoding with
the specific (and equally probable) unitaries $\{V_i\}$ and after the channel
action, as introduced in Lemma 1.
As an interpretation of this formula, note that the action of a  noisy channel
typically will increase the entropy of a given state, and therefore will decrease
the dense coding capacity of the original resource state.

In the next two sections we will study examples of channels and bipartite
states satisfying the condition (\ref{condition}), and evaluate explicitly the corresponding
super dense coding capacities.

\section{\emph{One-sided} $d$-dimensional Pauli channel}
 A $d$-dimensional Pauli channel \cite{generalizedpauli} that acts just on
Alice's side  is defined by
\begin{eqnarray}
\Lambda_a^{P}(\rho_i)=\sum_{m,n=0}^{d-1}q_{mn}(V_{mn}\otimes\id) \rho_i (V_{mn}^\dagger\otimes\id)\;,
\label{pauli-d-channel}
\end{eqnarray}
where $q_{mn}$ are probabilities (i.e. $q_{mn}\geq 0$ and $\sum_{mn}q_{mn}=1$).
The operators $V_{mn}$, defined in  (\ref{1}) with a slightly different
notation for the indices, can be expressed as

\begin{eqnarray}
V_{mn}=\sum_{k=0}^{d-1}\exp \left({\frac{2i\pi kn}{d}}\right)\ket{k}\bra{k+m(\mathrm{mod}\,
{d})}\;.
\label{vmn}
\end{eqnarray}
They satisfy $\tr V_{mn} = d\delta_{m0} \delta_{n0} $ and
$V_{mn} V_{mn}^\dagger=\id$, and have the properties

\begin{eqnarray}
V_{mn}V_{\tilde{m}\tilde{n}}=\exp \left({\frac{2i\pi \tilde{n}m }{d}}\right)V_{m+\tilde{m}(mod
\, {d}),n+\tilde{n}(mod\, {d})},
\label{groupproduct}
\end{eqnarray}

\begin{eqnarray}
\tr [V_{mn} V_{\tilde{m}\tilde{n}}^\dagger] = d\delta_{m\tilde{m}} \delta_{n\tilde{n}},
\label{vvdagger}
\end{eqnarray}

\begin{eqnarray}
 V_{mn} V_{\tilde{m}\tilde{n}}
=\exp\left({\frac{2i\pi(\tilde{n}m-n\tilde{m})}{d}}\right) V_{\tilde{m}\tilde{n}}V_{mn}.
\label{vv}
\end{eqnarray}
As the Kraus operators of one-sided Pauli channel (\ref{pauli-d-channel}) are unitary
it is a unital channel.

\subsection{Bell states}

A Bell state in $d\times d$ dimensions is defined as $\ket{\psi_{00}}= \frac{1}{\sqrt d}\sum_{j=0}^{d-1}\ket{j}\otimes \ket{j}$.
The set of the other maximally entangled Bell states is then denoted by  $\ket{\psi_{mn}}=(V_{mn}\otimes\id)\ket{\psi_{00}} $, for $m,n=0,1,...,d-1$.
We will show that for a Bell state shared between Alice and Bob, and 
with a \emph{one-sided} $d$-dimensional Pauli channel, the condition (\ref{condition}) is fulfilled.
We will first prove the following Lemma.

\textbf{Lemma 3.} Let us define
$\pi_{mn}:=(V_{mn}U\otimes\id)\rho_{00}(U^\dagger V_{mn}^\dagger\otimes\id)$, where
  $U$ is a unitary operator, $\rho_{00}=\ket{\psi_{00}}\bra{\psi_{00}}$ and
$V_{mn}$ is defined in (\ref{vmn}).
�For $m\not=\tilde{m}$,$n\not=\tilde{n}$,
\begin{eqnarray}
\pi_{mn}\pi_{\tilde{m}\tilde{n}}=0
\label{orto}
\end{eqnarray}
holds.\newline
\textbf{Proof.}\newline
In Appendix B  we show that $\rho_{00}(U^\dagger V_{mn}^\dagger V_{\tilde{m}\tilde{n}}U 
\otimes\id)\rho_{00}=0$ �for $m\not=\tilde{m}$,$n\not=\tilde{n}$.
Hence, 
\begin{eqnarray}
\pi_{mn}\pi_{\tilde{m}\tilde{n}}=(V_{mn}U\otimes\id)\underbrace{\rho_{00}(U^\dagger V_{mn}^\dagger V_{\tilde{m}\tilde{n}}U \otimes\id)\rho_{00}}_{0}� (U^\dagger V_{\tilde{m}\tilde{n}}^\dagger\otimes\id)=0\nonumber
\end{eqnarray}
\hfill{$\Box$}\newline
By using the orthogonality property (\ref{orto}) and the purity of the
density operators $\pi_{mn}$, we can write

   \begin{eqnarray}
 S(\Lambda_a^P(\tau))&=&S\left(\Lambda_a^P\left((U\otimes\id)\rho_{00}(U^\dagger\otimes\id)\right)\right)\nonumber\\
&=& S\left(\sum_{m,n=0}^{d-1}q_{mn}\underbrace{(V_{mn}U\otimes\id)\rho_{00} (U^\dagger V_{mn}^\dagger\otimes\id)}_{:=\pi_{mn}}\right)\nonumber\\
&=& 
H(\{q_{mn}\})\;,
\label{entropy-channel-bell}
\end{eqnarray}
where $H(\{q_{mn}\})=- \sum_{m,n} q_{mn} \log q_{mn} $ is the Shannon entropy.
We note that the von Neumann entropy $S(\Lambda_a^P(\tau))$ is independent of
the unitary encoding $U$. Consequently, for a \emph{one-sided} $d$-dimensional
Pauli channel with an initial  Bell state, the condition
(\ref{condition}) is satisfied. The super dense coding capacity
(\ref{capacity}) for an initial Bell state and a one-sided
Pauli channel in $d$ dimensions takes the  form

\begin{eqnarray}
C_{\mathrm{Bell}}^{\mathrm{one-sided\, P_d}}&=&S(\frac{\id}{d}\otimes\rho_b)-H(\{q_{mn}\})
=\log d^2-H(\{q_{mn}\})
\label{DCcapacityd-qubitcha}
\end{eqnarray}
for $m,n=0,1,...,d-1$.  Using (\ref{Cnoiseless}) we notice that the super dense coding capacity of a $d\times d$-dimensional Bell state
in the  noiseless case is given by $\log d^2$.
Thus,  in the presence of a one-sided Pauli channel
the super dense coding capacity is reduced by the amount $H(\{q_{mn}\})$
with respect to the noiseless case - i.e. the channel noise is simply subtracted
from the super dense coding capacity with noiseless channels.

Notice that the same capacity is achieved also for any maximally
entangled state, i.e. for any $\ket{\psi}=U_a\otimes U_b\ket{\psi_{00}}$.
Actually, Lemma 3 still holds in this case and therefore also the
derivation of the capacity (\ref{DCcapacityd-qubitcha}).


\subsection{Werner states}

We will now evaluate the super dense coding capacity for an input Werner
state $\rho_W=\frac{1-\eta}{d^2}\id + \eta\rho_{00}$ with $0\leq \eta \leq 1$.
The Werner state $\rho_W$ in the presence of a one-sided $d$-dimensional Pauli
channel provides another example of states and channels that satisfy
(\ref{condition}).\newline
 Using (\ref{entropy-channel-bell}), $\{q_{mn}\}$ is the set of eigenvalues of $ \Lambda_a^P\left[(U\otimes\id)\rho_{00}(U^\dagger\otimes\id)\right]$. The Pauli channel is a linear and unital map. Expressing the identity matrix $\id$ in a suitable basis,  we arrive at
\begin{eqnarray}
&&S\left(\Lambda_a^P\left((U\otimes\id)
\rho_{W}(U^\dagger\otimes\id)\right)\right)=S\left(\eta \Lambda_a^P\left[(U\otimes\id)
\rho_{00}(U^\dagger\otimes\id)\right]+\frac{1-\eta}{d^2}\id \right)\nonumber\\
&&=S\left(  {\rm diag}\left(\eta q_{00}+\frac{1-\eta}{d^2} ,...,\eta q_{d-1, d-1}+\frac{1-\eta}{d^2}\right) \right)\nonumber\\
&&=H\left( \{\eta q_{mn}+\frac{1-\eta}{d^2}\}\right).
\label{werner-paulichannel}
\end{eqnarray}
From (\ref{werner-paulichannel}) it is apparent that the output channel entropy is independent of the unitary encoding. Consequently, the super dense coding capacity, according to (\ref{capacity}), is given by

\begin{eqnarray}
C_{\mathrm{Werner}}^{\mathrm{one-sided\, P_d}}=\log d^2-H(\{\frac{1-\eta}{d^2}+\eta q_{mn}\}).
\end{eqnarray}
The above capacity is also achieved by any other state with the form $U_a\otimes U_b \rho_W U^\dagger_a\otimes U^\dagger_b$.

\section{Two-sided $d$-dimensional depolarizing channel.}

In  (\ref {pauli-d-channel}) we introduced the concept of a \emph{one-sided} $d$-dimensional Pauli channel. A \emph{two-sided} \emph{d}-dimensional Pauli channel is then defined by

\begin{eqnarray}
\Lambda_{ab}^{\mathrm{P}}(\rho_i)=\sum_{m,n,\tilde{m},\tilde{n}=0}^{d-1}
  q_{mn}q_{\tilde{m}\tilde{n}} (V_{mn}\otimes V_{\tilde{m}\tilde{n}})\rho_i (V_{mn}^\dagger\otimes V_{\tilde{m}\tilde{n}}^\dagger).
\label{qmn}
\end{eqnarray}
The \emph{d}-dimensional depolarizing channel is a special case of a \emph{d}-dimensional Pauli channel, with probability parameters

\begin{equation}
\label{cases}
q_{mn}=\cases{1-p+\frac p d^2,& $ m=n=0$\\
\frac p d^2 ,&  $\mathrm{otherwise}.$\\}
\end{equation}
for the noise parameter $p$, with $0\leq p \leq 1$, and $m,n=0,...,d-1$.

In the following Lemma we make the statement that the von Neumann 
entropy of a state that was  sent through the two-sided 
depolarizing channel is independent of any local unitary transformations
that were performed before the action of the channel.

\textbf{Lemma 4.} Let $\Lambda_{ab}^{\mathrm{dep}}$ denote a two-sided
\emph{d}-dimensional depolarizing channel. For a state $\rho$ and bilateral
unitary operator $U_a\otimes U_b$, we have
 \begin{eqnarray}
S\left(\Lambda_{ab}^{\mathrm{dep}} \left(\left(U_a\otimes U_b\right)\rho
(U_a^\dagger\otimes U_b^\dagger )\right)\right)=
S(\Lambda_{ab}^{\mathrm{dep}}(\rho)).
\label{depo-unitaryinvariant}
\end{eqnarray}

{\bf Proof}:
Considering
$\Lambda_{a}^{\mathrm{dep}}$  and $\Lambda_{b}^{\mathrm{dep}}$ to be the
\emph{d}-dimensional  depolarizing channels that act on Alice's
and Bob's system, respectively, it is straightforward to verify that

\begin{eqnarray}
\Lambda_{a}^{\mathrm{dep}}(\lambda_i)
=(1-p)\lambda_i \;,
\label{appendix0}
\end{eqnarray}
(where $\lambda_i$ are as before the generators of $SU(d)$),
and analogously for Bob's system.

Using the decomposition (\ref{rho}) for $\rho$ and the following
relation (proved in the Appendix A):

\begin{eqnarray}
\Lambda_{a}^{\mathrm{dep}}(U_a\lambda_iU_a^\dagger)
=(1-p) U_a \lambda_i U_a^\dagger\;,
\label{appendix}
\end{eqnarray}

it is then easy to prove the following covariance property of the channel:

\begin{eqnarray}
&&\Lambda_{ab}^{\mathrm{dep}} \left((U_a\otimes U_b)\rho
(U_a^\dagger\otimes U_b^\dagger)\right)
=(U_a\otimes U_b)\left[\Lambda_{ab}^{\mathrm{\mathrm{dep}}}(\rho)\right]
(U_a^\dagger\otimes U_b^\dagger).
\end{eqnarray}
Since the von Neumann entropy is invariant under unitary transformations,
the proof of Lemma 4 is complete.
\hfill{$\Box$}

As a consequence of Lemma 4 we can conclude that
for a two-sided $d$-dimensional depolarizing channel the entropy for a given
initial state $\rho$ is independent of the unitary encoding, namely

\begin{eqnarray}
S\left(\Lambda_{ab}^{\mathrm{dep}} \left(\left(U\otimes\id\right)\rho\left(U^\dagger\otimes\id\right)\right)\right)=S\left(\Lambda_{ab}^{\mathrm{dep}}(\rho)\right).
\label{38}
\end{eqnarray}
 Therefore, (\ref{condition}) holds and, according to
(\ref{capacity}), the super dense coding capacity for a given general resource state
$\rho$, with a two-sided $d$-dimensional depolarizing channel is  given by

\begin{eqnarray}
C^{\mathrm{two-sided\, dep_d}}(\rho)&=&S\left(\frac{\id}{d} \otimes\Lambda_b^{\mathrm{dep}}\left(\rho_b\right)\right)-S\left(\Lambda_{ab}^{\mathrm{dep}}\left(\rho\right)\right)\nonumber\\
&=& \log d +S\left(\Lambda_b^{\mathrm{dep}}\left(\rho_b\right)\right)-S\left(\Lambda_{ab}^{\mathrm{dep}}\left(\rho\right)\right).
\label{DC-2usedepo-any-state}
\end{eqnarray}

Notice that since Lemma 4 holds for any local unitary
$U_a\otimes U_b$,
the capacity (\ref{DC-2usedepo-any-state}) depends only on the degree
of entanglement of the input state $\rho$. In other words, all input states
with the same degree of entanglement have the same super dense coding capacity.

Comparing the above expression (\ref{DC-2usedepo-any-state}) with the one for the noiseless
case, given by $C=\log d+S(\rho_b)-S(\rho)$, one realizes that in the case of two-sided noise
the channel that affects Bob's subsystem enters twice, both in the von Neumann entropies
for the local and the global density matrix.

\subsection {Super dense coding capacity and optimal initial state 
}

In  (\ref{DC-2usedepo-any-state}) we obtained the super dense coding
capacity of an arbitrary given initial resource state $\rho$ for the \emph{two-sided} $d$-dimensional
depolarizing channel. In this subsection we perform the optimization of the
super dense coding capacity over the initial state of two qubits for the
\emph{two-sided} 2-dimensional depolarizing channel. Thus, we derive
the optimal value of the super dense coding capacity, if Alice and Bob have a 
depolarizing channel available for the transfer of 2-dimensional quantum states and can choose
the initial resource state.

A pure state of two qubits $\ket{\vartheta_\alpha}$ can be written in the Schmidt
bases $\{\ket{u_i}\},\{\ket{v_i}\}$ as $\ket{\vartheta_\alpha}= \sqrt{1-\alpha}
\ket{u_1v_1}+\sqrt{\alpha}\ket{u_2v_2} $ with $0 \leq \alpha \leq 1/2 $.
Two local unitaries $V_a$ and $V_b$ convert the computational bases to the
Schmidt
bases. Therefore, $\ket{\vartheta_\alpha}$ in computational bases can be
written as  $\ket{\vartheta_\alpha}=V_a \otimes V_b (\sqrt{1-\alpha}
\ket{00}+\sqrt{\alpha}\ket{11})$. In (\ref{depo-unitaryinvariant}) we
showed that the output von Neumann entropy of the two-sided depolarizing channel
is invariant under previous local unitary transformations. Therefore
$\ket{\vartheta_\alpha}$ and $\ket{\varphi_\alpha}=\sqrt{1-\alpha} \ket{00}
+\sqrt{\alpha}\ket{11}$ lead to the same dense coding capacity.
We can thus parametrize a pure initial state as a function of a single real
parameter, namely as the state $\ket{\varphi_\alpha}$, and follow the approach of
Ref. \cite{dep}.
The super dense coding capacity (\ref{DC-2usedepo-any-state}) of a pure state of two
qubits as a function of $\alpha $ and the noise
parameter $p$ is given by

\begin{eqnarray}
C^{\mathrm{two-sided\, dep_2}}_{\alpha} \left(\ket{\varphi_\alpha}\bra{\varphi_\alpha}\right)&=&1 - \xi_1 \log \xi_1
- \xi_2 \log\xi_2 \nonumber\\
&+& 
\gamma_1  \log \gamma_1 + \gamma_2 \log \gamma_2 +2 \gamma_3 \log \gamma_3\ ,
\label{capacity-alpha-p}
\end{eqnarray}
where  $\gamma_i$ (with $i=1,2,3,4$) are the eigenvalues of 
$\Lambda_{ab}^{\mathrm{dep}}\left(\ket{\varphi_\alpha}\bra{\varphi_\alpha}\right) $ and 
$\xi_s$ (with $s=1,2$) are the eigenvalues of $\Lambda_b^{\mathrm{dep}}\left(\rho_{b,\alpha} \right)$,  
where $\rho_{b,\alpha}=\tr_a(\ket{\varphi_\alpha}\bra{\varphi_\alpha})$.
 The eigenvalues $\gamma_i$ and $\xi_s$ are explicitly given by

\begin{eqnarray}
\gamma_{1,2}=\frac{1}{2}\left(1-p (1 -\frac{p}{2} )\pm  (1-p)\sqrt{1-4p\alpha(2-p)(1-\alpha)}\right) \ ,
 \nonumber\\
\gamma_{3}=\gamma_4 = \frac{p}{2} (1 -\frac{p}{2} )\ , \nonumber\\
\xi_1=\alpha-p\alpha+\frac{p}{2}  \ ,           \nonumber\\
\xi_2=1-\alpha+p\alpha-\frac{p}{2}\ .
\end{eqnarray}

We can now maximize  expression  (\ref{capacity-alpha-p})  over the variable
$\alpha $, for a given noise parameter $p$, and find interesting results.
They 
 are illustrated in Figure 3, where we plot  the superdense coding capacity in 
(\ref{capacity-alpha-p}) as a function of the noise parameter $p$, for various values $\alpha$.
We find that there is a threshold value $p_t\approx 0.345$, where two curves cross each other:
for $0\leq p \leq 0.345$ the value  $\alpha=1/2 $ leads to the highest super dense
coding capacity, i.e. the optimal initial resource state is a Bell state. For
$p\geq 0.345$, the optimal choice is $\alpha = 0$, i.e. product states are best
for dense coding. As shown graphically in the close-up of Figure 3, the curves for
intermediate values of $\alpha$ are always lower than $\alpha=1/2 $ or $\alpha = 0$. In order to prove this claim, we also evaluated   $C^{\mathrm{two-sided\, dep_2}}_{\alpha=1/2}- C^{\mathrm{two-sided\, dep_2}}_{\alpha}$ in the range of $0\leq p \leq 0.345$  and  $C^{\mathrm{two-sided\, dep_2}}_{\alpha=0}- C^{\mathrm{two-sided\, dep_2}}_{\alpha}$ in the range of $0.345\leq p \leq 1$ as functions of the parameters $\alpha$ and $p$. We found that these two functions are positive or zero.
Thus, for pure initial states it is always best to either use maximally entangled states
or product states, depending on the noise level.

\vspace{1cm}
\begin{center}
\begin{tabular}{c  }
\includegraphics[width=11cm]{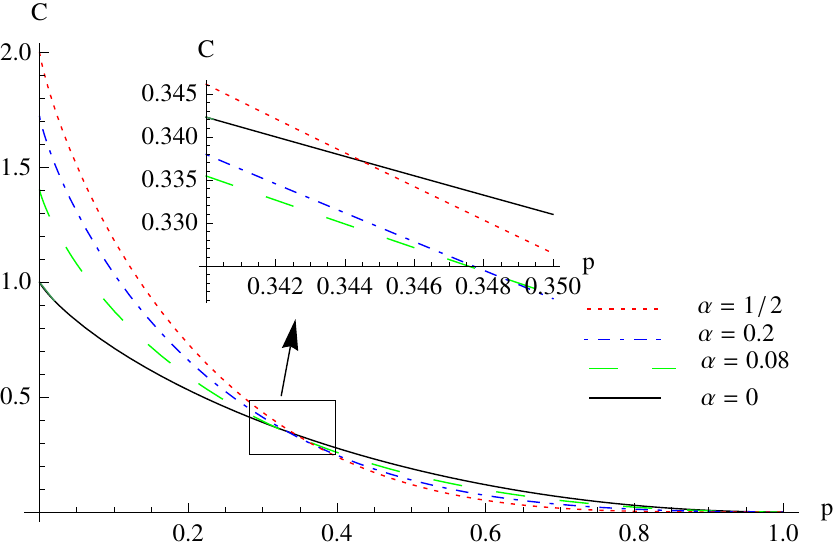}  \\
\end{tabular}
\end{center}
\vspace{0.5cm}

Figure  3. {\scriptsize The super dense coding capacity 
for the two-sided depolarizing channel in 2 dimensions,
$ C^{\mathrm{two-sided \, dep_2}}_{\alpha}$, as function of the noise parameter $p$,
 for  $\alpha=0$, $\alpha=0.08$, $\alpha=0.2$ and $\alpha=1/2$.  
For the definition of $\alpha$ see main text.
For $0\leq p \leq 0.345$ a Bell state, i.e. $\alpha = 1/2$, leads to the optimal 
capacity, while for  $0.345\leq p \leq 1$ the optimal initial state is a product state
($\alpha=0$). }

\vspace{1cm}

In the following we call the super dense coding capacity
of an initial  Bell state $\ket{\varphi_{1/2}}$ in the presence of a \emph{two-sided}
2-dimensional depolarizing channel 
$C^{\mathrm{two-sided\, dep_2}}_{\mathrm{Bell}}$. Using (\ref{capacity-alpha-p}) with $\alpha=1/2 $, this capacity is given by

\begin{eqnarray}
C^{\mathrm{two-sided\, dep_2}}_{\mathrm{Bell}}&=& 2+ \frac{1+3(1-p)^2}{4}\log  \frac{1+3(1-p)^2}{4} \nonumber\\
&+& 3 \frac{1-(1-p)^2}{4} \log \frac{1-(1-p)^2}{4}\ .
\label{DC-2usedepo}
\end{eqnarray}
The super dense coding capacity with an initial product state $\ket{\varphi_{0}}$ in the presence of a
\emph{two-sided} 2-dimensional depolarizing channel is denoted in the following as $C^{\mathrm{ch\, dep_2}}$. From (\ref{capacity-alpha-p}) with $\alpha=0$
it follows that

\begin{eqnarray}
C^{\mathrm{ch\, dep_2}}= 1+\frac{p}{2} \log\frac{p}{2}+ \frac{2-p}{2}\log\frac{2-p}{2}.
\label{channelcapacity}
\end{eqnarray}
Note that (\ref{channelcapacity}) is identical to  the classical channel capacity of
the depolarizing channel for qubits \cite{king}.


We now show that using mixed initial states as a resource 
cannot increase the super dense coding capacity,
i.e.
$\ket{\varphi_{1/2}}$ and $\ket{\varphi_{0}}$
are the optimal input states for the range of noise parameter
$0\leq p \leq 0.345$
and $0.345\leq p \leq 1$, respectively.
To show this claim we first write the super dense
coding capacity (\ref{DC-2usedepo-any-state}) in the form of the relative
entropy

\begin{eqnarray}
C^{\mathrm{two-sided\, dep_d}}(\rho)=
S(\Lambda_{ab}(\rho)\|\frac{\id}{d} \otimes\Lambda_{b}(\rho_b))\;.
\label{eq49}
\end{eqnarray}

Since any mixed state can be written as a convex combination of pure states
$\rho_k$, i.e. 
$\rho_{mix}=\sum_k p_k \rho_k$, and $\rho_{b,mix}=\tr_a(\rho_{mix})=
\sum_k p_k \rho_{b,k}$, we can  write

\begin{eqnarray}
 C_{\rho_{mix}}&=&S(\Lambda_{ab}(\rho_{mix})\|\tilde{\rho}) =
S( \Lambda_{ab}(\rho_{mix})\|\frac{\id}{d} \otimes\Lambda_{b}(\rho_{b,mix})) \nonumber\\
& =&S(\sum_k p_k \Lambda_{ab}(\rho_k)\|\sum_k p_k \frac{\id}{d}\otimes\Lambda_{b}(\rho_{b,k}))\nonumber\\
&\leq &\sum_k p_k S(\Lambda_{ab}(\rho_k)\| \frac{\id}{d}\otimes\Lambda_{b}(\rho_{b,k}))\;.
\end{eqnarray}

In the above inequality we have used the subadditivity of the relative entropy,
i.e. $S(\sum_i p_i r_i \| \sum_i q_i s_i)\leq \sum_i p_i S( r_i\|s_i)+
H(p_i\|q_i) $, where $H(\cdot\|\cdot)$ is the Shannon relative entropy,
defined as $H(p_i\|q_i)=\sum_i p_i \log \frac{p_i}{q_i} $ \cite{Nielsen}.
 We showed before that the super dense coding capacity of a pure state for
$0\leq p \leq 0.345$  is upper bounded by the super dense coding capacity of
a
Bell state $\ket{\varphi_{1/2}}$, and for  $0.345\leq p \leq 1$ it is upper
bounded by the product state $\ket{\varphi_{0}}$.
Remembering that $\rho_k$ is pure, and using (\ref{eq49}), we find that
 for $0\leq p \leq 0.345$

\begin{eqnarray}
 C_{\rho_{mix}} \leq \sum_k p_k S(\Lambda_{ab}(\rho_k)\| \frac{\id}{d}\otimes\Lambda_{b}(\rho_{b,k}))
\leq C^{\mathrm{two-sided \, dep_2}}\ , 
\end{eqnarray}

 and for $0.345\leq p \leq 1$

\begin{eqnarray}
 C_{\rho_{mix}} \leq \sum_k p_k S(\Lambda_{ab}(\rho_k)\| \frac{\id}{d}\otimes\Lambda_{b}(\rho_{b,k}))
\leq  C^{\mathrm{ch\, dep_2}}\ ,
\end{eqnarray}
which proves our claim.

It is interesting to note that the optimal
capacity for the two-sided qubit depolarizing channel is a non-differentiable
function of the noise parameter $p$, and that
the optimal states are either maximally entangled or separable. In other words,
there is a transition in the entanglement of the optimal input states
at the particular threshold value of the noise parameter
$p_t\approx 0.345$. Notice that a similar
transition behavior in the entanglement of the optimal input states for
transmission of classical information was found also for the qubit
depolarizing channel with correlated noise \cite{mem}.
It is interesting that in the present context the transition behavior arises
in a memoryless channel and is not related to correlations introduced via
the noise process.

\section{Super dense coding capacity versus channel capacity}

In this section, we consider the question of whether or not it is reasonable
in the presence of noise  to 
use the super dense coding protocol for the
transmission of classical information? To answer this question, we provide a comparison between  
the classical capacity of a 2-dimensional depolarizing channel and the super dense coding capacities of a
 \emph{one-sided} and \emph{two-sided} 2-dimensional depolarizing channel, for the resource of
an initial Bell state.
Since the depolarizing channel is a special form of a Pauli channel, according to 
(\ref{DCcapacityd-qubitcha}) the super dense coding capacity for a \emph{one-sided} 
2-dimensional depolarizing channel for an initially shared Bell state is
\begin{eqnarray}
C^{\mathrm{one-sided\, dep_2}}= 2+ \frac{4-3p}{4}\log  \frac{4-3p}{4} + 3 \frac{p}{4} \log \frac{p}{4}.
\label{DC-1usedepo}
\end{eqnarray}
The super dense coding capacity for
a \emph{two-sided} 2-dimensional depolarizing channel with  a Bell state as resource is
given in (\ref{DC-2usedepo}).
The classical capacity $C^{\mathrm{ch \, dep_2}}$ of the $2$-dimensional
depolarizing channel is achieved by an
ensemble of pure states belonging to an orthonormal basis,
say $\{\ket{0},\ket{1}\}$ at the channel input, with equal probability
$\frac{1}{2}$ and performing a complete von Neumann measurement in the same
basis over the channel output \cite{king}. Its expression is given explicitly
in  (\ref{channelcapacity}).

In Figure 4, we plot $C^{\mathrm{one-sided\, dep_2}}$,
$C^{\mathrm{two-sided\, dep_2}}$, $C^{\mathrm{ch\, dep_2}}$, and $C=1$  in terms of the noise parameter $p$.
As we expect, the first three capacities
$C^{\mathrm{one-sided\, dep_2}}$, $C^{\mathrm{two-sided\, dep_2}}$ and $C^{\mathrm{ch\, dep_2}}$ decrease as the  noise increases.
As expected, the super dense coding capacity of a \emph{one-sided}
$2$-dimensional depolarizing channel
$C^{\mathrm{one-sided\, dep_2}}$ is greater than the classical
capacity $C^{\mathrm{ch\, dep_2}}$ for all values of $p$, as
the additional resource of entanglement is used in dense coding.
The comparison between $C^{\mathrm{two-sided\, dep_2}}$ and
$C^{\mathrm{ch\, dep_2}}$ illustrates that for $0.345\leq p \leq 1$
the $2$-dimensional depolarizing  channel capacity is greater than the super
dense coding capacity for a \emph{two-sided} $2$-dimensional depolarizing
channel. This suggests that for $0.345\leq p \leq 1$  Alice and Bob do not
win by sending classical information via a super dense coding protocol with
unitary encoding. 
For this regime, the noise degrades the entanglement too much to be useful.
Now we can  answer the question posed at the beginning of
this section: super dense coding is not always a
useful scheme for sending classical information in the presence of noise.

\begin{center}
\begin{tabular}{c}
\includegraphics[width=10cm]{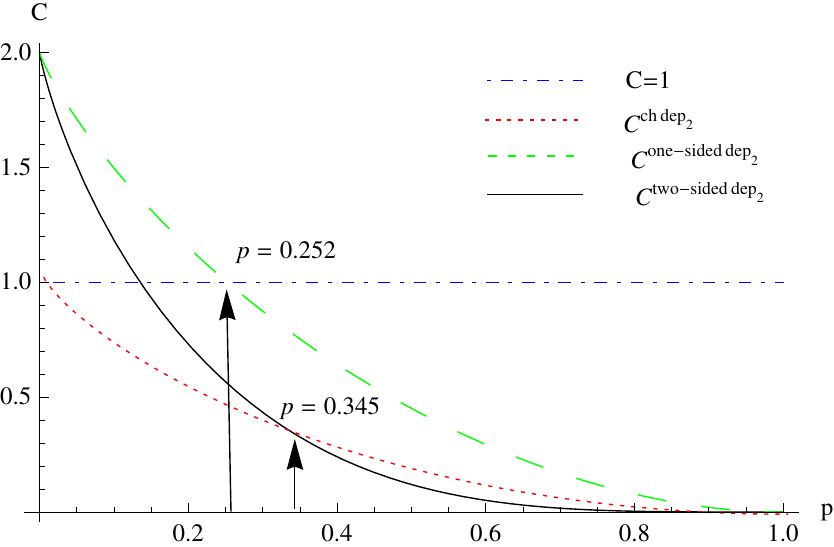}  \\
\end{tabular}
\end{center}

\vspace{0.5cm}

Figure  4. { \scriptsize The classical capacity  $ C^{\mathrm{ch\, dep_2}}$ of the
$2$-dimensional depolarizing
channel and the super dense coding
capacities for an initial Bell state in the presence of a \emph{one-sided} and
\emph{two-sided} 2-dimensional depolarizing channel,
$C^{\mathrm{one-sided\, dep_2}}$ and
$C^{\mathrm{two-sided\, dep_2}}$, respectively, as functions of the noise
parameter $p$.}

\vspace{1cm}

We notice also that $C^{\mathrm{one-sided\, dep_2}}$ corresponds to
the entanglement assisted capacity for the depolarizing channel \cite{ben99}.
According to (\ref{DC-1usedepo}) for $p=0.252$ the super dense coding
capacity for an initial  Bell state via the \emph{one-sided} 2-dimensional
depolarizing channel is equal to \emph{one}. The maximum information that can be
transmitted by two-dimensional systems without any source of entangled states
is $C=1$. That is, for $p=0.252$ the super dense coding capacity reaches the
classical limit, as can be seen in Figure 4.
It was shown in \cite{horodeski-teleportation} that the classical limit of the quantum teleportation 
protocol, when using a  Bell state and distributing one subsystem of it via a
depolarizing channel, is reached at $p=1/3$. In the absence of noise, quantum 
teleportation and super dense coding are two equivalent protocols \cite{werner}. 
According to our results  this is not true in the presence of noise, as we have shown explicitly for the depolarizing
channel: here, the quantum/classical boundary for super dense coding
occurs at a different noise
value than for quantum teleportation.

We point out that the expression (\ref{DCcapacityd-qubitcha})
for the dense coding capacity
of a Bell state provides a lower bound to the entanglement-assisted capacity
of a general Pauli channel.

\section{Non-unitary encoding for the
$d$-dimensional Pauli channel}
So far, we assumed that the encoding in the super dense coding protocol
is unitary.
In this section we consider the possibility of performing non-unitary encoding
on the initial state and discuss explicitly the case of the depolarizing channel.
Let us consider $\Gamma_i$ to be a completely positive trace preserving (CPTP) map. Alice applies the map $\Gamma_i$ on her side of the shared state $\rho$, thereby  encoding $\rho$ as $ \rho^i= [\Gamma_i \otimes \id] (\rho):=\Gamma_i(\rho)$. The super dense coding protocol with non-unitary encoding for noiseless channels has been discussed by M. Horodecki et al. \cite{unitaryoptimal1},
M. Horodecki and Piani \cite{CPTP},  and Winter
\cite{unitaryoptimal2}.
In this section  we introduce an upper bound on the Holevo quantity
for a \emph{two-sided} $d$-dimensional Pauli channel, and
show that this upper bound is  reachable by a pre-processing before unitary
encoding.
Our arguments follow a similar line as in \cite{CPTP}, 
where non-unitary encoding was studied for the case of noiseless channels.

\textbf{Lemma 5.} Let $\chi=S\left( \sum_i p_i \Lambda^P_{ab}(\rho_i)\right)-\sum_i p_i S
\left(\Lambda^P_{ab}\left(\rho_i\right)\right)$ be the Holevo quantity with  $ \rho_i=\Gamma_i(\rho)$ 
and let $\Lambda^P_{ab}(\rho)$ be a general \emph{two-sided} $d$-dimensional Pauli channel defined via

\begin{eqnarray}
\Lambda^P_{ab}(\rho)= \sum_{m,n,\tilde{m},\tilde{n}=0}^{d-1}
  q_{mn\tilde{m}\tilde{n}}(V_{mn}\otimes V_{\tilde{m}\tilde{n}})(\rho) (V_{mn}^\dagger\otimes V_{\tilde{m}\tilde{n}}^\dagger)
\label{map}
\end{eqnarray}
with $\sum_{m,n,\tilde{m},\tilde{n}=0}^{d-1}
  q_{mn\tilde{m}\tilde{n}}=1 $. Let $\Gamma_M(\cdot):=[\Gamma_M \otimes \id](\cdot)$  be the map 
that minimizes the von Neumann entropy  after application
of this map and  the channel $\Lambda_{ab}$ to
the initial state $\rho$, i.e. $\Gamma_M$ minimizes the expression
 $S\left(\Lambda^P_{ab}(\Gamma_M (\rho))\right)$.
 Then, the Holevo quantity $\chi$ is upper bounded by

\begin{eqnarray}
\chi \leq \log d + S\left(\Lambda^P_b(\rho^B)\right)- 
    S\left(\Lambda^P_{ab}(\Gamma_M (\rho))\right).
\label{upperbound}
\end{eqnarray}

\textbf{Proof:}  $\Gamma_M(\cdot)$ is a map that leads to the minimum of the entropy 
after applying it and the channel to the initial state $\rho$.
 Therefore,

\begin{eqnarray}
\chi&=&S\left( \sum_i p_i \Lambda^P_{ab}(\rho^i)\right)-\sum_i p_i S\left(\Lambda^P_{ab}\left(\rho^i\right)\right)\nonumber\\
&\leq& S\left( \sum_i p_i \Lambda^P_{ab}(\rho^i)\right)- S \left(\Lambda^P_{ab}(\Gamma_M (\rho))\right).\nonumber
\end{eqnarray}
Since the von Neumann entropy is subadditive
and since the maximum entropy of a $d$-dimensional system is $\log d $, we have
\begin{eqnarray}
\chi&\leq& \log d + S\left(\tr_a \left(\sum_i p_i \Lambda^P_{ab}(\rho^i)\right)\right) - S \left(\Lambda^P_{ab}\left(\Gamma_M (\rho)\right)\right). \nonumber
\end{eqnarray}
Now, since $\tr_a \sum_i p_i \Lambda^P_{ab}(\rho_i)=\Lambda^P_b(\rho_b)$ it follows that

\begin{eqnarray}
\chi &\leq& \log d + S\left(\Lambda^P_b(\rho_b)\right)-S \left(\Lambda^P_{ab}\left(\Gamma_M (\rho)\right)\right).\nonumber
\end{eqnarray}
\hfill{$\Box$}

 If the upper bound in (\ref{upperbound}) is achievable, then it is equal to 
the super dense coding capacity. We consider
the ensemble $\{{\tilde{ p_i},\tilde{\Gamma}_i(\rho)}\}$ with $ \tilde{p_i}=\frac{1}{d^2}$ and 
 $\tilde{\Gamma}_i(\rho)=(V_i \otimes\id) \Gamma_M (\rho)(V_i^\dagger \otimes\id)$,
 where $V_i$ is defined  in (\ref{1}).
We will show in the following that this ensemble achieves the upper bound
in (\ref{upperbound}). In other words, the optimal encoding consists of a fixed pre-processing
with $\Gamma_M$ and a subsequent unitary encoding. This is analogous to the case
of noiseless channels, for which the same statement was shown in
\cite{CPTP}.
The Holevo quantity of the ensemble $\{{\tilde{ p_i},\tilde{\Gamma}_i(\rho)}\}$ is

\begin{eqnarray}
\tilde{\chi}= S \left( \sum_i\frac{1}{d^2}\Lambda^P_{ab}\left(\tilde{\Gamma}_i(\rho)\right)\right)-\sum_i \frac{1}{d^2}  S\left[\Lambda^P_{ab}\left(\tilde{\Gamma}_i(\rho)\right)\right].
\label{optiensemble}
\end{eqnarray}

By using (\ref{averagenoiseless}) and noting that $\Gamma_M$ acts only on Alice's side,
and  by using Lemma 1-a), we find that the 
average of $ \Lambda^P_{ab}\left(\tilde{\Gamma}_i(\rho)\right)$,
i.e. the argument in the first term on the RHS of (\ref{optiensemble}),
is given by

\begin{eqnarray}
\sum_i\frac{1}{d^2}\Lambda^P_{ab}\left(\tilde{\Gamma}_i(\rho)\right)
=\frac{\id}{d} \otimes\Lambda^P_b({\rho_b}).
\label{average-state}
\end{eqnarray}

Furthermore, 
 the second term on the RHS of (\ref{optiensemble})
is given by

\begin{eqnarray}
&&\sum_i \frac{1}{d^2}  S\left(\Lambda^P_{ab}\left(\tilde{\Gamma}_i(\rho)\right)\right)=  \sum_i \frac{1}{d^2} S\left( \Lambda^P_{ab}\left( \left(V_i \otimes\id\right) {\Gamma_M}(\rho)\left(V_i^\dagger \otimes\id\right)\right)\right)\nonumber\\
&&=\frac{1}{d^2} \sum_{i} S\left((V_{i}\otimes \id ) \left[\sum_{m,n,\tilde{m},\tilde{n}=0}^{d-1}
q_{mn\tilde{m}\tilde{n}}  \left(V_{mn}\otimes V_{\tilde{m}\tilde{n}}\right) {\Gamma_M}(\rho)\left(V_{mn}^\dagger \otimes V_{\tilde{m}\tilde{n}}^\dagger \right)\right]\cdot\right.\nonumber\\
&&\left. \hspace{0.4cm}\cdot (V_{i}^\dagger\otimes \id)\right)\nonumber\\
&&= \frac{1}{d^2} \sum_{i}S\left[\Lambda^P_{ab}\left({\Gamma_M}(\rho)\right)\right]
= S\left[\Lambda^P_{ab}\left({\Gamma_M}(\rho)\right)\right]\,
\label{average-entropy}
\end{eqnarray}
where in the second line of the above equations we have 
inserted the action of the channel, defined in (\ref{map}), and we have 
used (\ref{vv}), from which it follows that $V_i$ and $V_{mn}$ commute up to a phase. 

Inserting (\ref{average-state}) and (\ref{average-entropy}) 
into (\ref{optiensemble}), one finds that 
the Holevo quantity $\tilde{\chi} $ is equal to  the upper bound given in (\ref{upperbound}). Consequently, the super dense coding capacity with non-unitary encoding is
\begin{eqnarray}
C=\log d + S\left(\Lambda^P_b(\rho_b)\right)-S \left[\Lambda^P_{ab}\left(\Gamma_M (\rho)\right)\right]\;.
\end{eqnarray}
Thus, we
have shown above for the case of a $d$-dimensional Pauli channel that
applying the appropriate pre-processing $\Gamma_M$ on the initial state $\rho$ before 
the unitary encoding $\{V_i\}$  may increase the super dense coding capacity,
with respect to only using unitary encoding.
Our results derived in  section 4 provide  an example where
pre-processing indeed leads to an improvement:
 Consider the case
of a \emph{two-sided} 2-dimensional depolarizing
channel for an initial  Bell state with  a noise parameter in the range
$0.345\leq p \leq 1$, see Figure 3.
To reach the optimal super dense coding capacity in this case, Alice applies a
measurement as a pre-processing, projecting the Bell state onto
$\ket{00}$ or $\ket{11}$; afterwards she applies the unitary encoding.
As we showed above, the super dense coding capacity for product states
is equal to the capacity of the
depolarizing channel, given in (\ref{channelcapacity}). Thus, in 
this case we  reach a higher super dense coding capacity than without pre-processing. The effect of pre-processing is illustrated in Figure 5, which is an excerpt of Figure 4.

\begin{center}
\begin{tabular}{c}
\includegraphics[width=10cm]{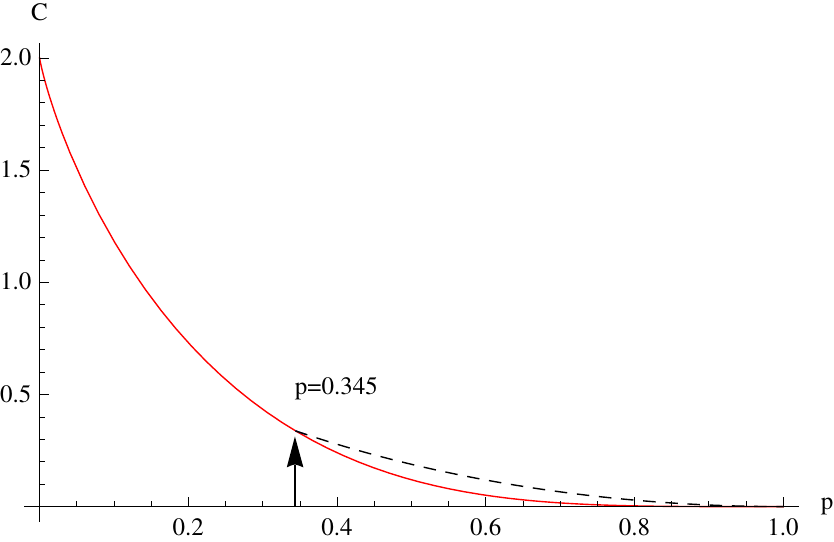}  \\
\end{tabular}
\end{center}
Figure 5. {\scriptsize The solid curve is the optimal super dense coding capacity with a Bell state in the presence of a \emph{two-sided} 2-dimensional depolarizing channel. The dashed line shows the improved super dense coding capacity by using a pre-processing on the Bell state in the range of 
$0.345\leq p \leq 1$.
}

\section{Conclusions}

In conclusion, we investigated the bipartite super dense coding protocol
in the presence of a unital noisy channel, which acts either only on Alice's
subsystem after encoding (\emph{one-sided} channel) or both on
Alice's and Bob's subsystems (\emph{two-sided} channel). For those cases where the von Neumann
entropy fulfills a specific condition, we derived the super dense coding
capacity. We showed that a \emph{one-sided} $d$-dimensional Pauli channel for
the resource of Bell
and Werner states fulfills the above mentioned condition on the von Neumann
entropy.
Our condition on the von Neumann entropy is also satisfied for
a  \emph{two-sided} $d$-dimensional depolarizing channel.
For these examples, we derived the explicit optimal super dense coding capacity,
as a function of the initial resource state.
When the initial state can be chosen, we found
for the case of a \emph{two-sided} $2$-dimensional depolarizing channel that the 
optimal initial resource state is either a Bell state or a product state, depending on the value of  noise parameter.

 We also compared the classical capacity of the 2-dimensional depolarizing channel  to the
super dense coding capacities for an initial Bell state
with a  \emph{one-sided} and  \emph{two-sided}
2-dimensional  depolarizing channel.
Our results showed that Alice and Bob may not win by sending classical
information via a super
dense coding protocol with unitary encoding, if there is too much noise. 
Comparing the critical noise parameters for the quantum/classical boundary,
we found that in the scenario of the depolarizing channel the protocols 
quantum teleportation and super dense coding are not equivalent, in the sense that they
do not have the same critical noise parameter.
 
Finally,   we discussed the super dense coding capacity with non-unitary encoding for  a \emph{two-sided}   $d$-dimensional  Pauli channel. We showed that the optimal strategy is to   apply  
a pre-processing before the unitary encoding. We gave an example of  super dense coding for an initial Bell state and a \emph{two-sided} 2-dimensional  depolarizing channel where
 pre-processing increases the  super dense coding capacity, as compared to only unitary encoding.

There are several open questions: how can the super dense coding capacity be
determined  for other channels and states than the ones that fulfil the specific
entropy condition? What is the influence of correlated noisy channels?
How does noise affect the multipartite super dense coding scenario? These
topics will be addressed in future work.
 
 \ack We are grateful for discussions
with Alexander Holevo, Barbara Kraus and Colin Wilmott. This work was
partially supported by the EU Integrated Project SCALA, the European Project CORNER  and Deutsche Forschungsgemeinschaft (DFG).


\newpage
\appendix
\section{ }

We give here a proof for (\ref{appendix}). We expand $U\lambda_iU^\dagger$ in terms of  $\{V_{mn}\}$. By using the fact that $ \lambda_i$ is traceless,
we have
\begin{eqnarray}
\Lambda_{a}^{dep}(U\lambda_iU^\dagger)&=& \Lambda_{a}^{dep}\left(\sum_{m,n\neq(0,0)}^{d-1}\gamma_{mn}V_{mn}\right)\nonumber .
\end{eqnarray}
Here, $\Lambda_{a}^{dep}(\cdot)$ is a linear map that is given by $\Lambda_{a}^{dep}(\cdot)= \sum_{\tilde{m},\tilde{n}=0}^{d-1} q_{\tilde{m}\tilde{n}}V_{\tilde{m}\tilde{n}}(\cdot)V_{\tilde{m}\tilde{n}}^\dagger $. Then we can write
\begin{eqnarray}
\Lambda_{a}^{dep}(U\lambda_iU^\dagger)&=&\sum_{m,n\neq(0,0)}^{d-1}\gamma_{mn} \Lambda_{a}^{dep}(V_{mn})\nonumber\\
&=&\sum_{\tilde{m},\tilde{n}=0}^{d-1}\sum_{m,n\neq(0,0)}^{d-1}\gamma_{mn}q_{\tilde{m}\tilde{n}}V_{\tilde{m}\tilde{n}} V_{mn} V_{\tilde{m}\tilde{n}}^\dagger \nonumber .
\end{eqnarray}

By using (\ref{vv}) and unitarity of $V_{mn}$, we have

\begin{eqnarray}
\Lambda_{a}^{dep}(U\lambda_iU^\dagger)&=&\sum_{\tilde{m},\tilde{n}=0}^{d-1}  \sum_{m,n\neq(0,0)}^{d-1}\gamma_{mn} q_{\tilde{m}\tilde{n}}\exp\left({\frac{2i\pi(n\tilde{m}-\tilde{n}m)}{d}}\right) V_{mn} \nonumber\\
&=&\sum_{m,n\neq(0,0)}^{d-1}\gamma_{mn} V_{mn}\sum_{\tilde{m},\tilde{n}=0}^{d-1}q_{\tilde{m}\tilde{n}}\exp\left({\frac{2i\pi(n\tilde{m}-\tilde{n}m)}{d}}\right).\nonumber
\end{eqnarray}

For $q_{\tilde{m}\tilde{n}}$ we replace the expression of (\ref{qmn})
and we then have

\begin{eqnarray}
\Lambda_{a}^{dep}(U\lambda_iU^\dagger)&=&\sum_{m,n\neq(0,0)}^{d-1}\gamma_{mn} V_{mn}\left(1-p+ \frac{p}{d^2}\underbrace{\sum_{\tilde{m},\tilde{n}=0}^{d-1}\exp\left({\frac{2i\pi(n\tilde{m}-\tilde{n}m)}{d}}\right)}_{\delta_{0,m}\delta_{0,n}}\right)\nonumber\\
&=& (1-p)\sum_{m,n\neq(0,0)}^{d-1}\gamma_{mn} V_{mn}
= (1-p) U \lambda_i U^\dagger\nonumber
\end{eqnarray}
 Therefore, (\ref{appendix}) is proved.


\section{}

In Lemma 3  we need to prove that $\rho_{00}(U^\dagger V_{mn}^\dagger V_{\tilde{m}\tilde{n}}U \otimes\id)\rho_{00}=0$. We here show that 
$\bra{\psi_{00}}(U^\dagger V_{mn}^\dagger V_{\tilde{m}\tilde{n}}U \otimes\id)\ket{\psi_{00}}=0$,
from which the previous statement follows. Due to (\ref{vvdagger}) for $m\not=\tilde{m}$ and $n\not=\tilde{n}$ the expression� $ V_{mn}^\dagger V_{\tilde{m}\tilde{n}}$ is traceless and $\{V_{jk}\}_{j,k=0}^{d-1}$ form a complete set. We� can thus expand $ V_{mn}^\dagger V_{\tilde{m}\tilde{n}}=\sum_{(j,k)\not=(0,0)} \beta_{jk}V_{jk} $ with expansion coefficients $\beta_{jk}$. Therefore,
\begin{eqnarray}
&&\bra{\psi_{00}}(U^\dagger V_{mn}^\dagger V_{\tilde{m}\tilde{n}}U \otimes\id)\ket{\psi_{00}}=\sum_{(j,k)\not=(0,0)}\beta_{jk}�� \bra{\psi_{00}}(U^\dagger V_{jk}U \otimes\id)\ket{\psi_{00}}\nonumber\\
&&= \frac{1}{d}\sum_{(j,k)\not=(0,0)}\sum_{m,n=0}^{d-1}\beta_{jk}\bra{mm}(U^\dagger V_{jk}U\otimes\id) \ket{nn}\nonumber\\
&&=\frac{1}{d}\sum_{(j,k)\not=(0,0)}\sum_{m,n=0}^{d-1}\beta_{jk}\bra{m}U^\dagger V_{jk}U\ket{n} \bracket{m}{n}\nonumber\\
&&=\frac{1}{d}\sum_{(j,k)\not=(0,0)}\beta_{jk}\tr[U^\dagger V_{jk}U]
= \frac{1}{d}\sum_{(j,k)\not=(0,0)}\beta_{jk}\tr [V_{jk}]=0. \nonumber
\end{eqnarray}
Since $\rho_{00}= \ket{\psi_{00}}\bra{\psi_{00}}$, we arrive at 
\begin{eqnarray}
�\rho_{00}(U^\dagger V_{mn}^\dagger V_{\tilde{m}\tilde{n}}U \otimes\id)\rho_{00}=0\ ,
\end{eqnarray}
which completes the proof.

\cleardoublepage


\end{document}